\documentclass[numberedappendix,natbib209]{emulateapj}

\usepackage{epsfig}
\usepackage{longtable}
\usepackage{natbib}

\begin{document}

\title{Discovery of a Rich Cluster at \lowercase{z} = 1.63 using the Rest-Frame 1.6$\micron$ ``Stellar Bump Sequence" Method}
\author{Adam Muzzin\altaffilmark{1}, Gillian Wilson\altaffilmark{2}, Ricardo Demarco\altaffilmark{3}, Chris Lidman\altaffilmark{4}, Julie Nantais\altaffilmark{3}, Henk Hoekstra\altaffilmark{1}, H. K. C. Yee\altaffilmark{5}, Alessandro Rettura\altaffilmark{6}}


\altaffiltext{1}{Leiden Observatory, Leiden University, PO Box 9513,
  2300 RA Leiden, The Netherlands}
\altaffiltext{2}{Department of Physics and Astronomy,
University of California, Riverside, CA 92521}
\altaffiltext{3}{Department of Astronomy, Universidad de Concepcion, Casilla 160-C, Concepcion, Chile}
\altaffiltext{4}{Australian Astronomical Observatory, P.O. Box 296, Epping NSW 1710, Australia}
\altaffiltext{5}{Department. of Astronomy \& Astrophysics, University
  of Toronto, 50 St. George St., Toronto, Ontario, Canada, M5S 3H4}
\altaffiltext{6}{Department of Astronomy, California Institute of Technology, MC 249-17, Pasadena, CA, 91125, USA}
\begin{abstract}
We present a new two-color algorithm, the ``Stellar Bump Sequence" (SBS), that is optimized for robustly identifying candidate high-redshift galaxy clusters in combined wide-field optical and mid-infrared (MIR) data.  The SBS algorithm is a fusion of the well-tested cluster red-sequence method of Gladders \& Yee (2000) with the MIR 3.6$\micron$ - 4.5$\micron$ cluster detection method developed by Papovich (2008).  As with the cluster red-sequence method, the SBS identifies candidate overdensities within 3.6$\micron$ - 4.5$\micron$ color slices, which are the equivalent of a rest-frame 1.6$\micron$ stellar bump ``red-sequence".  In addition to employing the MIR colors of galaxies, the SBS algorithm incorporates an optical/MIR (z$^{\prime}$ - 3.6$\micron$) color cut.  This cut effectively eliminates foreground 0.2 $< z <$ 0.4 galaxies which have 3.6$\micron$ - 4.5$\micron$ colors that are similarly red as $z >$ 1.0 galaxies and add noise when searching for high-redshift galaxy overdensities.   We demonstrate using the $z \sim$ 1 GCLASS cluster sample that similar to the red sequence, the stellar bump sequence appears to be a ubiquitous feature of high-redshift clusters, and that within that sample the color of the stellar bump sequence increases monotonically with redshift and provides photometric redshifts accurate to $\Delta$$z$ = 0.05.  We apply the SBS method in the XMM-LSS SWIRE field and show that it robustly recovers the majority of confirmed optical, MIR, and X-ray-selected clusters at $z > 1.0$ in that field.  Lastly, we present confirmation of SpARCS J022427-032354 at $z =$ 1.63, a new cluster detected with the method and confirmed with 12 high-confidence spectroscopic redshifts obtained using FORS2 on the VLT.  We conclude with a discussion of future prospects for using the algorithm. 

\end{abstract}
\section{Introduction}
Massive galaxy clusters are important astrophysical objects.  As the largest virialized structures in the universe, their space density, mass function, and clustering provide robust constraints on cosmological models \citep[e.g.,][]{Edge1990,Eke1998,Bahcall2003,Gladders2007,Vikhlinin2009,Rozo2010,Sehgal2011,Benson2011}.  As the highest-density regions in the universe, they are also important cosmic laboratories for studying how environment affects the evolution of galaxies \citep[e.g.,][and numerous others]{Dressler1980,Balogh1999,Poggianti1999,Poggianti2006,Patel2009,Patel2009b,Vulcani2010,Wetzel2011b,Muzzin2012}.  For these reasons there has always been strong motivation to discover larger samples of clusters, increasingly more massive clusters, and naturally, clusters at the highest-possible redshifts.
\newline\indent
Substantial progress on all these fronts has been made in the last decade, particularly in finding clusters to the highest redshifts.  Ten years ago only a few clusters at $z >$ 1 had been discovered and confirmed \citep{Stanford1997,Rosati1999}.  Five years ago only a few clusters at $z > 1.4$ had been confirmed \citep{Stanford2005,Stanford2006}.  Now, using a multitude of techniques, the number of confirmed clusters above $z > $ 1 is approaching $\sim$ 100, and the number of confirmed clusters at $z > 1.4$ is now $>$ 10.  
\newline\indent
The progress in finding high-redshift clusters has occurred at an extremely rapid pace, but perhaps what has been most astonishing is that new high-redshift clusters are now being discovered routinely using different techniques and observations at a wide range of wavelengths.  In previous decades searches for the highest-redshift clusters were almost completely dominated by X-ray surveys \citep[e.g.,][]{Gioia1990,Rosati1998}, whose superior sensitivity over wide areas allowed them to robustly discover massive clusters with relative ease.  In recent years the advent of wide-field near- and mid-infrared imaging (hereafter NIR and MIR, respectively), as well as millimeter-wavelength surveys for clusters using the Sunyaev-Zel'dovich effect (hereafter SZE) has made these methods competitive with the latest generation of X-ray surveys.
\newline\indent
The current large sample of $z >$ 1 clusters comes from X-ray surveys \citep[e.g.,][]{Mullis2005,Stanford2006,Bremer2006,Pacaud2007,Henry2010,Fassbender2011a,Fassbender2011b,Nastasi2011,Santos2011}, space-based MIR surveys \citep[e.g.,][]{Stanford2005,Eisenhardt2008,Muzzin2009a,Wilson2009,Papovich2010,Demarco2010,Brodwin2011,Stanford2012,Zeimann2012}, ground-based NIR surveys \citep[e.g.,][]{Gobat2011,Spitler2012}, millimeter SZE surveys \citep[e.g.,][]{Brodwin2010,Foley2011,Stalder2012}, as well as narrow-band and MIR searches around high-redshift radio galaxies \citep[e.g.,][]{Venemans2007,Galametz2010,Galametz2012}.
\newline\indent
Naturally, this array of detection methods has produced a heterogeneous sample of high-redshift clusters.  While this may create some difficulty when comparing the properties of clusters and cluster galaxies between various samples, there are also distinct advantages to having samples of clusters selected in diverse ways.  For example, SZE and X-ray surveys detect clusters based on their baryonic mass in the form of hot gas, and therefore preferentially select halos with higher gas densities and higher gas fractions.  Optical/IR surveys detect clusters based on their stellar mass, and therefore preferentially select halos with higher stellar-mass fractions.  If the relative fraction of baryons in the form of gas and stars varies substantially from halo to halo \citep[e.g.,][]{Lin2003,Dai2010}, then these techniques are complementary -- indeed almost necessary -- for finding mass-complete samples of clusters.
\newline\indent
For technical reasons, X-ray and SZE cluster surveys have historically covered large volumes searching for the most massive systems, whereas optical/IR surveys have covered smaller volumes and have primarily discovered low-mass systems.  Optical/IR surveys over large volumes are advantageous because they can discover complementary samples of the rare massive clusters typically selected in SZE and X-ray surveys.  It is challenging to obtain optical/IR data in multiple filters over wide areas; however, \cite{Gladders2000,Gladders2005} demonstrated that clusters can be selected in wide-area surveys with a low contamination rate with just two photometric bands up to $z$ = 1 using the red-sequence of cluster early-type galaxies as a high-contrast signpost for clusters.  This red-sequence method was adapted into the MIR for the SpARCS survey \citep{Muzzin2009a,Wilson2009}, where the 50 deg$^2$ $Spitzer$ SWIRE fields \citep{Lonsdale2003} were observed in the z$^{\prime}$-band and clusters to $z \sim$ 1.4 were detected using a z$^{\prime}$ - 3.6$\micron$ color to identify galaxy overdensities.  
\newline\indent
SpARCS was a logical extension of the red-sequence method into the MIR in order to search for clusters at higher redshift than optical surveys; however, the technique does require deep optical data to compliment the MIR data, a considerable investment in observing time.  \cite{Papovich2008} proposed a novel method for selecting high-redshift clusters using MIR data alone, thus avoiding the need for additional deep optical data.  \cite{Papovich2008} demonstrated that high-redshift galaxies tend to have red 3.6$\micron$ - 4.5$\micron$ colors, and that because of the negative k-correction in these bands, even distant galaxies are bright enough to be detected in relatively shallow MIR imaging.  He devised a color cut (3.6$\micron$ - 4.5$\micron$ $>$ -0.1 AB) that allowed the identification of numerous cluster candidates in the SWIRE field at $z >$ 1.3, and this approach led to the identification of the highest-redshift known cluster at that time \citep{Papovich2010}.  
\newline\indent
In this paper we present the ``Stellar Bump Sequence" (SBS) which is an improved algorithm for efficiently detecting high-redshift clusters in combined medium-deep optical/MIR imaging data.  The method is a two color (three filter) algorithm based on combining the red-sequence cluster detection method used in SpARCS with the MIR method as developed by \cite{Papovich2008, Papovich2010}.  In this paper we show that the hybrid of the two methods improves the reliability of candidate overdensities and also provides good-quality photometric redshifts for all cluster candidates.  
We apply this new technique to the XMM-LSS SWIRE field, and show that it recovers the majority of known clusters at $z > 1$ in that field, with good estimates of their photometric redshift.  Lastly, we present a new, rich cluster spectroscopically confirmed at $z = 1.63$ that was discovered using the method.
\newline\indent
The layout of this paper is as follows.  In $\S$2 we present the SBS algorithm and demonstrate that the stellar bump sequence is clearly seen in spectroscopically-confirmed clusters at $z \sim$ 1.  In $\S$3 we apply the algorithm to the XMM-LSS field and show that it recovers the majority of known clusters at $z >$ 1 in that field.  In $\S$4 we present the confirmed cluster J022427-032354 at $z = $1.63, a demonstration of the effectiveness of the algorithm for selecting high-redshift clusters.  We conclude with a discussion in $\S$5.  Throughout this paper we assume a $\Omega_{\Lambda}$ = 0.7, $\Omega_{m}$ = 0.3, and H$_{0}$ = 70 km s$^{-1}$ Mpc$^{-1}$ cosmology.  All magnitudes are in the AB system.  
\begin{figure*}
\plotone{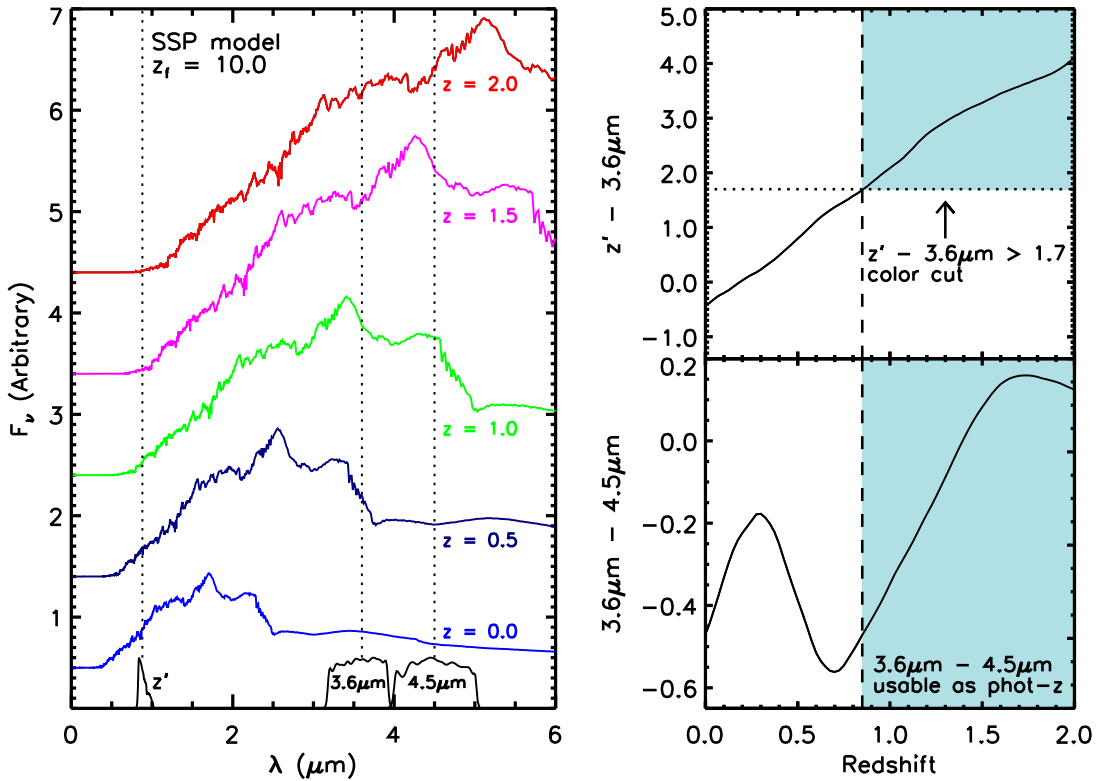}
\caption{\footnotesize Left Panel: The evolving SED of an SSP with a $z_{f}$ = 10 generated with the BC03 models as it would be observed at various redshifts.  Filter response curves for the z$^{\prime}$, 3.6$\micron$, and 4.5$\micron$ bands are shown below and the central wavelengths are marked with the dotted vertical lines.  Top Right Panel: The z$^{\prime}$ - 3.6$\micron$ color as a function of redshift for the SEDs in the left panel.  The z$^{\prime}$ - 3.6$\micron$ color increases monotonically with redshift and hence provides an estimate of the photo-z for an SSP.  Bottom Right Panel:  The 3.6$\micron$ - 4.5$\micron$ color as a function of redshift for the SEDs in the left panel.  Due to the more complex SED shape in the rest-frame NIR, the 3.6$\micron$ - 4.5$\micron$ color does not uniquely map onto redshift; however, if a z$^{\prime}$ - 3.6$\micron$ $>$ 1.7 color cut is applied to cull low-redshift galaxies (see top right panel), the 3.6$\micron$ - 4.5$\micron$ color becomes a viable estimate of the photometric redshift as those filters move through the rest-frame 1.6$\micron$ stellar bump feature.}
\end{figure*}
\section{The Stellar Bump Sequence Algorithm}
\subsection{Motivation}
\indent
Photometric redshifts for galaxies are typically determined by fitting the observed colors of galaxies with a range of templates for galaxy spectra.  The features in galaxy spectral energy distributions (SEDs) that provide the tightest constraints on their photometric redshift are those that are the sharpest, particularly the Lyman break at 912\AA, the Balmer break at $\sim$ 3800\AA, and the 4000\AA~break in old stellar populations.  An additional strong feature is the so-called ``stellar bump" feature at rest-frame $\sim$ 1.6$\micron$ which is caused by the minimum opacity of the H$^{-}$ ion at this wavelength \citep[e.g.,][]{John1988}.   This results in a maximum of the flux density at 1.6$\micron$ for galaxies whose ages are $>$ 10 Myr \citep[]{Sawicki2002}.  As an illustration of the stellar bump feature, in Figure 1 we plot the SEDs of a single-burst simple stellar population (SSP) with solar metallicity formed at $z$ = 10 as it would be seen at $z =$ 0.5, 1.0, 1.5, and 2.0.  The SEDs have been generated using the \cite{Bruzual2003} models.  We note that we have used an SSP as a fiducial model for the purpose of illustration, but that any stellar population with an age $>$ 10 Myr also shows the 1.6$\micron$ feature \citep[see e.g.,][]{Sawicki2002,Papovich2008}.  The 1.6$\micron$ stellar bump feature can be seen clearly in the SEDs in Figure 1, and its prominence resulted in it being discussed prior to the launch of $Spitzer$ as a potentially powerful photometric redshift indicator \citep{Simpson1999,Sawicki2002}.  
\newline\indent
In practice, the stellar bump does provide constraints on the photometric redshifts of galaxies; however, these tend to be much weaker than those obtained from the bluer features such as the 4000\AA~break.  In the lower right panel of Figure 1 we plot the 3.6$\micron$ - 4.5$\micron$ color of this model as a function of redshift.  The observed color maps uniquely onto redshift within the redshift interval 0.7 $< z < $1.7; however, the complete range in color is small ($\sim$ 0.7 mag), making the color only a modest-precision estimator of photometric redshift.  In the upper right panel of Figure 1 we plot the observed z$^{\prime}$ - 3.6$\micron$ color of the model as a function of redshift.  This color spans the 4000\AA~break at $z >$ 1, and in contrast to the 3.6$\micron$ - 4.5$\micron$ color, increases monotonically with redshift at all redshifts, and has a substantial range ($\sim$ 4 mag).  These properties make it a more powerful tool for identifying high-redshift cluster candidates and estimating their redshifts.  
\newline\indent
In the SpARCS survey \citep{Muzzin2009a,Wilson2009} we used the z$^{\prime}$ - 3.6$\micron$ colors of galaxies to detect clusters up to $z \sim$ 1.4.  However, as Figure 1 shows, at $z >$ 1.4 the z$^{\prime}$ - 3.6$\micron$ colors of the model become extremely red (z$^{\prime}$ - 3.6$\micron$ $>$ 3.0).  This implies that z$^{\prime}$-band imaging that is substantially deeper than the MIR data are required to identify candidate high-redshift clusters with this method.  This is approximately the limiting depth of the SpARCS z$^{\prime}$ data (z$^{\prime}$ $<$ 24.2), which makes it challenging to select clusters at $z >$ 1.4 robustly with the red-sequence method.
\begin{figure*}
\plotone{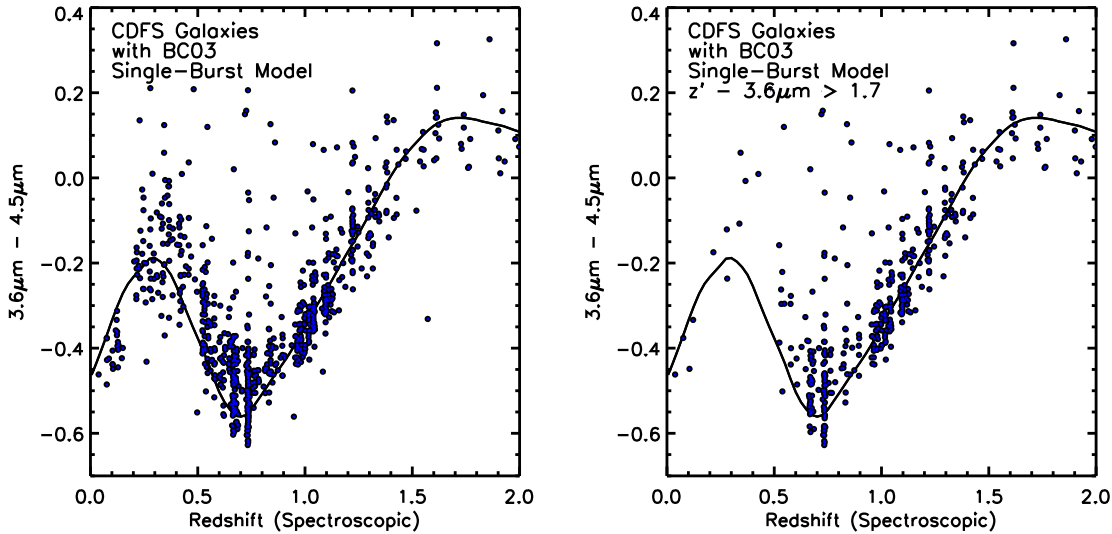}
\caption{\footnotesize Left Panel: The observed 3.6$\micron$ - 4.5$\micron$ color as a function of spectroscopic redshift for all galaxies in the FIREWORKS catalog of Wuyts et al (2008) with $3.6\micron$ $<$ 22.0.  An SSP with a $z_{f}$ = 10 generated using the BC03 models is overplotted for reference and is a reasonable description of the colors of most galaxies.  This occurs because the 3.6$\micron$ - 4.5$\micron$ colors of galaxies are fairly insensitive to their star formation history.  Galaxies that lie considerably above the curve are likely to have a warm dust component heated by an AGN.  Right Panel:  Same as left panel, but with galaxies with z$^{\prime}$ - 3.6$\micron$ $>$ 1.7 plotted.  The color cut excludes most galaxies at $z < $ 0.7 with red 3.6$\micron$ - 4.5$\micron$ colors and shows that for the remaining galaxies the 3.6$\micron$ - 4.5$\micron$ color can be used as a photo-z accurate to $\pm$ 0.15 for an individual galaxy.}
\end{figure*}
\newline\indent
\cite{Papovich2008} showed that provided only clusters at $z > $ 1.3 were considered, the 3.6$\micron$ - 4.5$\micron$ color of galaxies is sufficient to select high-redshift overdensities of galaxies.  This novel idea allowed him to take full advantage of the deep SWIRE MIR data (a 5$\sigma$ depth of $\sim$ 0.5 L$^{*}$ at $z \sim$ 2), to select distant cluster candidates.   \cite{Papovich2008} searched for overdensities of galaxies in position space that passed a single color cut of 3.6$\micron$ - 4.5$\micron$ $>$ -0.1 in the SWIRE fields and discovered hundreds of $z > $ 1.3 cluster candidates including the confirmed cluster ClG J0218.3-5010 \citep{Papovich2010}.  
\newline\indent
This new method has been an important step forward for pushing MIR cluster detection to the highest redshifts, and ClG J0218.3-5010 has been an important object for studying galaxy evolution in high-redshift, high-density environments \citep[e.g..][]{Tran2010,Papovich2011,Quadri2012,Rudnick2012,Tadaki2012}.  Still, a single cluster almost certainly cannot be representative of the full population of clusters at $z >$ 1.4 and much larger samples of clusters selected in the MIR that are mass-complete and have a high purity level would be extremely valuable.  As of yet, it is unclear whether searching for clusters with only the \cite{Papovich2008} 3.6$\micron$ - 4.5$\micron$ color cut is sufficiently robust to produce such samples.  
\newline\indent
There are several disadvantages to the color-cut method as compared to red-sequence selection \citep[e.g.,][]{Muzzin2009a,Wilson2009} or multi-band photometric redshift selection \citep[e.g.,][]{Eisenhardt2008}.  
Specifically, the fact that 3.6$\micron$ - 4.5$\micron$ does not uniquely map onto redshift means that galaxies at redshifts 0.2 $< z <$ 0.4 have similar colors to those at $z >$ 1.0 and add noise when selecting high-redshift clusters.  \cite{Papovich2008} argued that this contamination is minimized provided the MIR color cut was red enough (3.6$\micron$ - 4.5$\micron$ $>$ -0.1); however, as we show in $\S$ 2.2.1, even with such a red color cut the contamination is considerable, particularly at bright magnitudes.  
\newline\indent
The second issue is the lack of the use of additional color information beyond the initial 3.6$\micron$ - 4.5$\micron$ cut.  Such selection compresses all structures at 1.3 $< z <$ 2.0 along the line-of-sight.  This type of selection is similar to that adopted by the early-generation optical cluster surveys such as the Abell catalog.  Those cluster catalogs suffer from low purity levels because numerous cluster candidates are in fact line-of-sight projections of low-mass groups.  The projection effect problem in early optical cluster surveys has now been remedied by employing selection algorithms such as the red-sequence method, which detects clusters in narrow color ``slices" \citep{Gladders2000,Gladders2005}, hence reducing the probability of line-of-sight projections.  Color slices are not used in the \cite{Papovich2008} method, which suggests that a cluster sample selected using that method may suffer similar projection effect issues as in the early optical cluster surveys.
\newline\indent
The last issue with the MIR color-cut approach is that it does not provide information on the photometric redshift of cluster candidates (other than the cluster should be at $z > $ 1.3).  Whether they are used for cosmological purposes, or for studying the galaxy population, the photometric redshift distribution for a cluster sample is a key quantity.  Furthermore, given the current lack of known clusters at 1.5 $< z <$ 2.0, being able to differentiate cluster candidates at higher redshift compared to lower redshift is a distinct advantage for prioritizing followup observations.
\newline\indent
Here we outline some modifications that we have made to the MIR cluster detection method proposed by \cite{Papovich2008} that address some of these issues.  The two main changes are the use of an additional optical/MIR color cut, and the selection of clusters within multiple MIR color slices.  This effectively combines the MIR color approach of \cite{Papovich2008} with the well-tested cluster red sequence method of \cite{Gladders2000,Gladders2005}.  We show that the optical/MIR color cut substantially reduces the contamination of high-redshift galaxies by low-redshift galaxies and is so efficient at removing interlopers that the MIR selection method can be used down to much lower redshift ($z \sim$ 0.85).  Similar to the red-sequence approach, the 3.6$\micron$ - 4.5$\micron$ color slice method reduces the number of line-of-sight projections, and also provides an estimate of the photometric redshift for each cluster candidate.
\begin{figure*}
\plotone{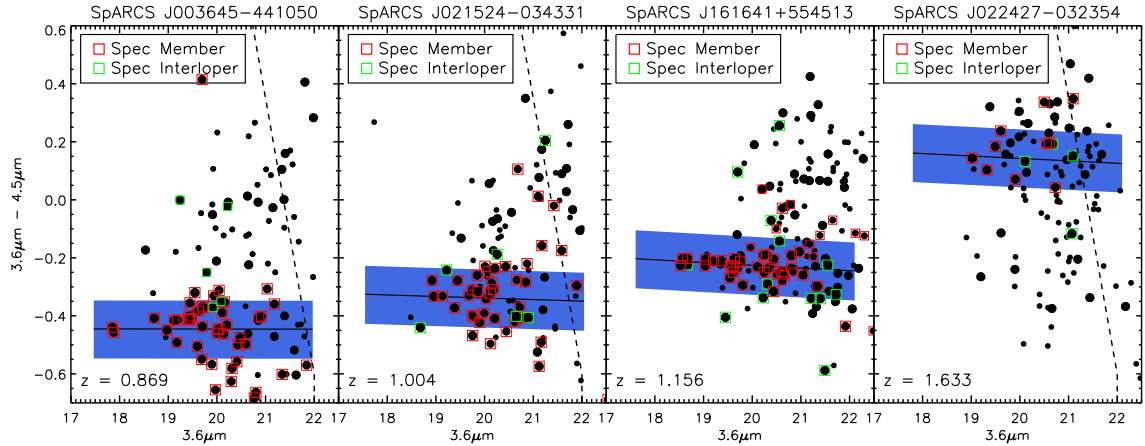}
\caption{\footnotesize The 3.6$\micron$ - 4.5$\micron$ vs. 3.6$\micron$ color-magnitude diagrams for four SpARCS clusters.  The three left-most clusters were discovered with the z$^{\prime}$ - 3.6$\micron$ red-sequence method and have extensive spectroscopy from the GCLASS survey.  Galaxies within R $<$ 0.5 Mpc and 0.5 Mpc $<$ R $<$ 1.0 Mpc are plotted as large and small symbols, respectively.  Galaxies that have spectroscopic redshifts concurrent with being clusters members are shown as red squares, those with spectroscopic redshifts inconsistent with being clusters members are shown as green squares.  The cluster members belong to a fairly tight (scatter $\sim$ 0.1 mag) sequence in color-magnitude space that becomes monotonically redder with increasing redshift.  Similar to the red-sequence method, this ``stellar-bump sequence" can be used as a signature to identify clusters in wide-field imaging.  The solid lines and blue shaded regions are models for the stellar-bump sequence determined using BC03 models.  The 3.6$\micron$ - 4.5$\micron$ CMD for SpARCS J022427-032354, which was identified from its stellar-bump sequence is shown in the right-most panel and is substantially redder than the lower-redshift SpARCS clusters.}
\end{figure*}
\subsection{Modifications to the Papovich Method}
\subsubsection{z$^{\prime}$ - 3.6$\micron$ color cut}
In the left panel of Figure 2 we plot the observed 3.6$\micron$ - 4.5$\micron$ color vs. spectroscopic redshift for galaxies with 3.6$\micron$ $<$ 22.0 mag (the 5$\sigma$ depth of the SWIRE survey) in the GOODS-South FIREWORKS catalog of \cite{Wuyts2008}.  Overplotted as the solid line is the color of the single-burst model shown in Figure 1.  The single-burst model is a reasonable description of the galaxy population at all redshifts, primarily because the 3.6$\micron$ - 4.5$\micron$ colors of galaxies are insensitive to star formation history.  
\newline\indent
Figure 2 shows that while there is a good correspondence between 3.6$\micron$ - 4.5$\micron$ color and redshift for galaxies at $z >$ 0.7, there is a significant population of 0.2 $< z <$ 0.4 galaxies with colors similar to those at $z >$ 1.  \cite{Papovich2008} argued that using a color cut of 3.6$\micron$ - 4.5$\micron$ $>$ -0.1 excludes most of the 0.2 $< z <$ 0.4 population.  The number of $z >$ 1.3 galaxies in the FIREWORKS spectroscopic sample with 3.6$\micron$ - 4.5$\micron$ $>$ -0.1 and 3.6$\micron$ $<$ 22.0 mag is 82.  The number of $z <$ 0.4 galaxies matching the same color/magnitude criteria is 23.  Although the Papovich MIR color criteria does eliminate some low redshift interlopers, clearly the interloper fraction is still non-negligible, $\sim$ 20\%.  We note that 20\% is formally an upper limit for the contamination because the spectroscopy in GOODS-South is more complete at $z <$ 0.4 than at $z >$ 1.3.  Still, the overall spectroscopic completeness is quite good \citep[see][]{Wuyts2008}, particularly at these bright magnitudes, so the estimate of 20\% should be indicative of the actual interloper fraction.
\newline\indent
In the right panel of Figure 2 the FIREWORKS galaxies are plotted again, this time excluding galaxies with $z^{\prime}$ - 3.6$\micron$ $<$ 1.7.  This removes the majority of the 0.2 $< z <$ 0.4 population without excluding the $z >$ 1.3 population.  The color cut removes 20/23 of the low-redshift interlopers ($\sim$ 85\%), but only removes 5 galaxies at $z >$ 1.3 ($\sim$ 6\%).  Therefore, with the additional $z^{\prime}$ - 3.6$\micron$ $>$ 1.7 criteria, the interloper fraction form galaxies at $z <$ 0.4 is improved by a factor of five to an impressively low $\sim$ 4\%.  
\newline\indent
Clearly a significant reduction in projection effects from low-redshift interlopers can be made by including a simple optical/MIR color cut in addition to the Papovich 3.6$\micron$ - 4.5$\micron$ color selection.  \cite{Papovich2008} also showed similar reductions in the interloper fraction if a simple R-band magnitude cut was employed (R $<$ 22.5 mag).  Indeed, making a color cut does not even require particularly deep optical data.  The SpARCS z$^{\prime}$ data in the SWIRE fields is approximately 2.2 mag deeper than the 3.6$\micron$ data, deeper than is needed to make the $z^{\prime}$ - 3.6$\micron$ $>$ 1.7 cut.  This implies that future surveys employing the SBS selection could use $\sim$ 3$\times$ less integration time than SpARCS in the z$^{\prime}$-band, which would be only 35 minutes per pixel on a 4m-class telescope.  
\subsubsection{Color Slices vs. a Color Cut}
Within Figure 2, several overdensities in spectroscopic redshift are clearly visible.  These overdensities have mean 3.6$\micron$ - 4.5$\micron$ colors consistent with the single-burst model and fairly small dispersions around this color ($\sim$ 0.1 mag).  This suggests that using slices in 3.6$\micron$ - 4.5$\micron$ color would be an effective way to minimize projection effects when selecting high-redshift clusters.  This approach is identical to that taken for red-sequence selection, other than it uses very different rest-frame wavelengths which bracket the 1.6$\micron$ stellar bump feature rather than the 4000\AA~break.  
\newline\indent
To further demonstrate the well-behaved nature of the MIR colors of cluster galaxies, in Figure 3 we plot the 3.6$\micron$ - 4.5$\micron$ colors vs. 3.6$\micron$ flux for galaxies in the fields of three clusters at $z =$ 0.869, 1.004, and 1.156 with extensive spectroscopy from the GCLASS survey \citep[see][]{Muzzin2012}.  Confirmed spectroscopic members are shown as red squares and confirmed non-members are shown as green squares.  Similar to the overdensities in FIREWORKS, the colors of the cluster members are nearly identical.  It appears that similar to the red sequence of early-types, there is a ``stellar bump sequence" for galaxies in high-redshift clusters that increases monotonically with redshift (within this redshift window).  
\newline\indent
The stellar bump sequence also appears to be quite tight.  Overplotted in each panel is a model for the sequence (see $\S$ 2.3) with a shaded region that represents a scatter of 0.1 mag.  This is consistent with the rms scatter around the model for galaxies with 3.6$\micron$ $<$ 21.0 seen in the two lowest redshift clusters which are 0.093 and 0.078 mag, respectively (rejecting $>$3$\sigma$ outliers).  The data for those clusters is from the SWIRE survey that has 150s integration time per pixel and the median photometric error is 0.05 mag.  The $z =$ 1.156 cluster \citep[see also][]{Demarco2010} has data from a GTO followup program (PI: Fazio/Wilson) and has an exposure time of 2040s per pixel.  The scatter in the stellar bump sequence for that cluster is substantially smaller, 0.047 mag (again rejecting $>$3$\sigma$ outliers).  The median photometric error in the deeper data is 0.02 mag.  Subtracting the photometric scatters from the measured scatters in quadrature for all three clusters suggests that the stellar bump sequence has an impressively small intrinsic scatter of 0.04 -- 0.07 mag.  
\newline\indent
Figure 3 shows that selecting clusters in narrow slices of 3.6$\micron$ - 4.5$\micron$ color should be an effective way to minimize projection effects.  The sequences in Figure 3 are spaced such that there should be little overlap within the photometric scatter for clusters $\delta$$z$ $>$ 0.15 apart in redshift space.  This is not as good as the precision from red-sequence selection where sequences typically only overlap within the photometric scatter for clusters with $\delta$$z$ $<$ 0.05 \citep{Gladders2005,Muzzin2008}; however it is a significant improvement over no color selection whatsoever.  
\newline\indent
Lastly, we note that the solid lines shown in Figure 3 are not fits to the data, but are \cite{Bruzual2003} models for the 3.6$\micron$ - 4.5$\micron$ sequence at the redshift of the cluster assuming a single burst with a $z_{f}$ = 10 .  These are remarkably good descriptions of the data and show that the color slice method should also return a reasonable estimate of the photometric redshift for the cluster (we discuss this further in $\S$ 5).
\begin{figure*}
\plotone{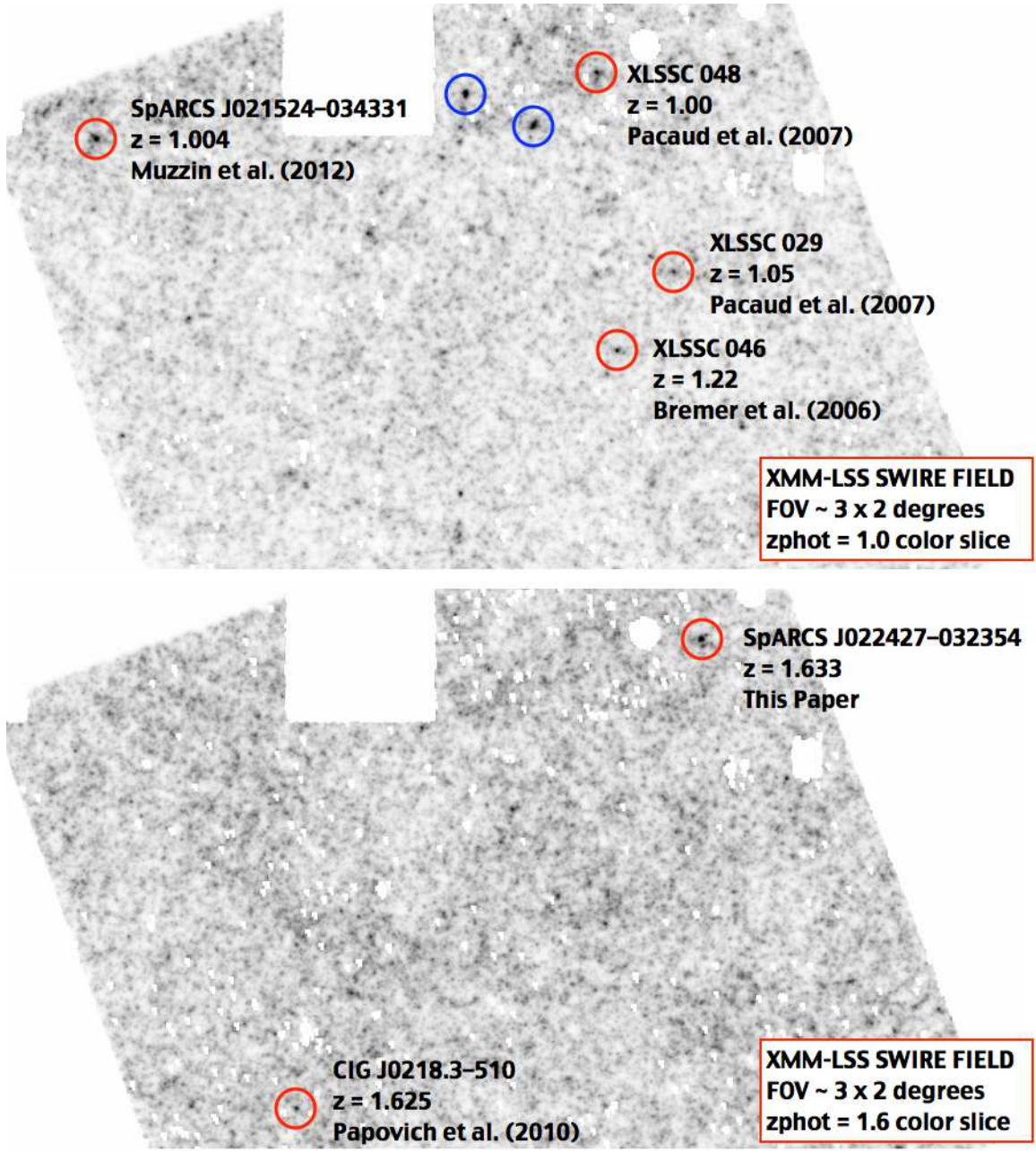}
\caption{\footnotesize Top Panel: Smoothed density map of galaxies in the XMM-LSS SWIRE field with z$^{\prime}$ - 3.6$\micron$ $>$ 1.7 and -0.5 $<$ 3.6$\micron$ - 4.5$\micron$ $<$ -0.3.  The field of view on the sky is $\sim$ 3 $\times$ 2 degrees, north is up, east is to the right.  The locations of spectroscopically-confirmed clusters discovered with X-ray observations (XLSSC clusters) and the red-sequence method (SpARCS J021524-034331) are shown with red circles.  Those clusters are clearly associated with overdensities in the map and this shows that the 2-color SBS-method recovers clusters discovered with other search techniques.  Shown in blue are two new cluster candidates discovered with the technique (see text).  Bottom Panel: Same as top panel but for galaxies in a redder color slice, 0.0 $<$ 3.6$\micron$ - 4.5$\micron$ $<$ 0.2.  The cluster SpARCS J022427-032355 is clearly visible in the upper right and is the most significant overdensity in the map.  The cluster discovered by Papovich et al. (2010) using the SWIRE observations is also visible in the bottom left and is also one of the more significant overdensities in this color slice.}
\end{figure*}
\subsection{The SBS Algorithm}
The SBS algorithm is nearly identical to the red-sequence detection algorithm proposed by \cite{Gladders2000} and applied in \cite{Gladders2005} and \cite{Muzzin2008} and we refer to those papers for a detailed step-by-step description of the method.  In brief, the method is the following.  Firstly, several SSP models for the apparent 3.6$\micron$ - 4.5$\micron$ colors of galaxies as a function of redshift is determined using the \cite{Bruzual2003} models.  These models are normalized using the $U$ - $V$, $V$ - $I$ and $J$ - $K$ colors of the red sequence in the Coma cluster \citep{Bower1992}.  These models are then used to construct color slices analagous to those plotted in Figure 3.  For each color slice the probability that a given galaxy in the survey area belongs to that slice is calculated, the result being a probability weight for each galaxy, in each slice.  In this process we only calculated weights for galaxies that pass the z$^{\prime}$ - 3.6$\micron$ $>$ 1.7 color-cut.
\newline\indent
Probability maps are then created for each slice by placing galaxies on a pixelated grid of the survey field in their observed positions, weighted by their probabilities.  The maps are then smoothed with a kernel with a physical size of 0.5 Mpc FWHM, and object detection is performed to identify sources above the background.  Sources in the probability map are then flagged as candidate galaxy clusters.  Sources are typically found in several color slices and a final catalog of candidates is made by connecting detections between slices and assigning candidates a photometric redshift based on the slice where they have the most significant detection.  An example of SBS probability maps in the XMM-LSS SWIRE field in two different color slices is shown in Figure 4.  
\begin{figure*}
\plotone{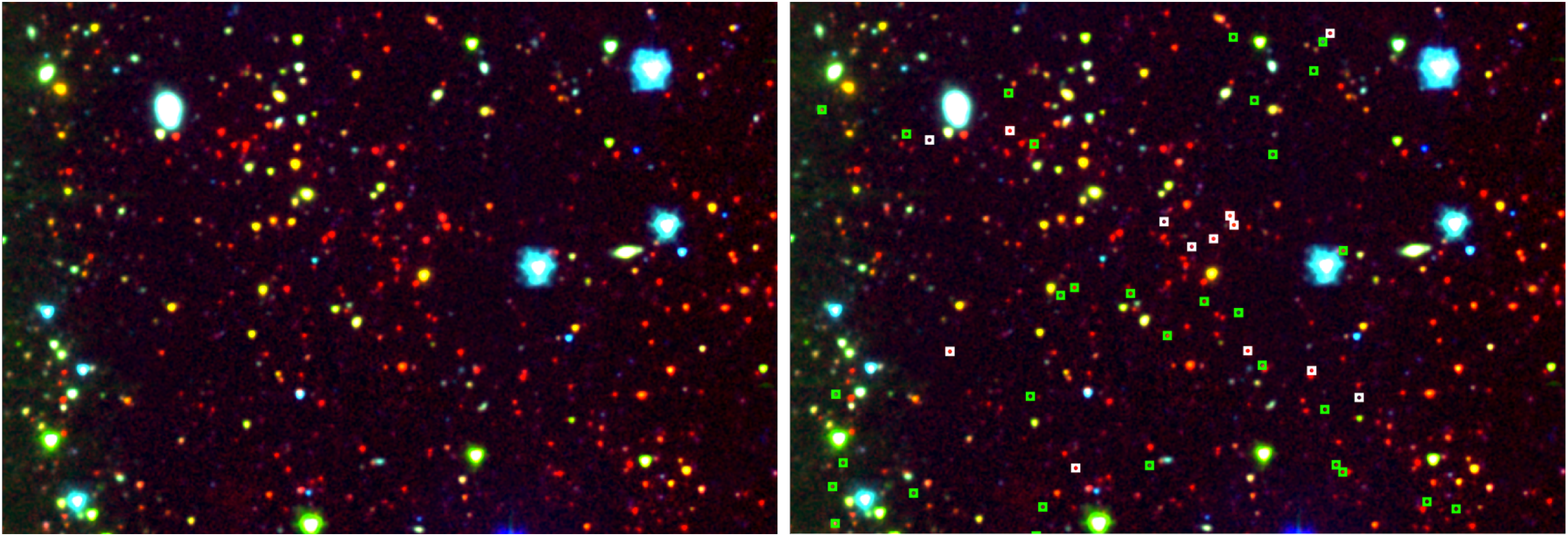}
\caption{\footnotesize Left Panel: gz$^{\prime}$3.6$\micron$ color composite of the cluster SpARCS J022427-032354.  The field-of-view is 7$^{\prime}$ $\times$ 5$^{\prime}$. Right Panel: Same as left panel but with spectroscopically-confirmed cluster members shown in white and spectroscopically-confirmed foreground/background galaxies shown in green.}
\end{figure*}
\section{Application to the XMM-LSS SWIRE Field}
\indent
The XMM-LSS field contains X-ray data from the $XMM-Newton$ satellite taken as part of the XMM-LSS cluster survey \citep{Pierre2006}.  The field has been searched extensively for high-redshift clusters using a variety of techniques, making it an ideal testbed for the SBS algorithm.  There are four spectroscopically-confirmed $z >$ 1 X-ray-selected clusters from the XMM-LSS: XLSSC005, XLSSC029, XLSSC048, and XLSSC046 \citep{Bremer2006,Pacaud2007,Adami2011}.  There is one confirmed cluster from the SpARCS survey (SpARCS J021524-034331) discovered with the z$^{\prime}$ - 3.6$\micron$ red-sequence method \citep{Muzzin2012}.  The field also contains the spectroscopically-confirmed cluster CIG J0218.3-510 at $z =$ 1.62 discovered by \cite{Papovich2010} using the MIR color-cut method.  
\newline\indent
The field contains four channel IRAC imaging from the SWIRE survey \citep[see][]{Lonsdale2003} covering a total area of 9.4 deg$^2$. There is also deep z$^{\prime}$ data taken as part of the CFHTLS-Wide survey, as well as an extended region of z$^{\prime}$ data coverage taken as part of a CFHT followup program (PI: Pierre, ID: 2006BF11).  We use these datasets to apply the SBS method and search for clusters at 0.85 $< z <$ 2.0
\newline\indent
In Figure 4 we show the probability maps generated using the SBS method.  The top panel shows the color slice corresponding to a $z_{photo}$ = 1.0, the bottom panel shows the same field but generated with galaxies in the color slice $z_{photo}$ = 1.6.  In both panels the width of the slice in color corresponds to a width in redshift of $\delta$$z$ $\sim$ $\pm$ 0.1.  In the $z_{photo}$ = 1.0 slice there are several overdensities that are clearly visible.  Those that are associated with the confirmed $z >$ 1 clusters are circled in red.  When object detection is run, the SBS algorithm recovers 3/4 X-ray selected clusters, as well as the red-sequence selected SpARCS cluster.  There appears to be no overdensity associated with the 4th X-ray cluster (XLSSC005), although we note that for this cluster the spectroscopic redshift is based on only 2 members and therefore its spectroscopic redshift may not be secure \citep{Pacaud2007}.
\newline\indent
Figure 4 shows that the SBS selection appears to work well, recovering most of the high-confidence clusters at $z \sim$ 1 selected with other methods.  As shown by the blue circles in Figure 4, the method has also identified two other $z \sim$ 1 overdensities with even higher significance than any of the confirmed clusters.  These two candidates, SpARCS J022158-034000 and SpARCS J022056-033348 have estimated photometric redshifts of 1.08 and 1.10, respectively and are not in the latest XMM-LSS catalogs \citep{Pacaud2007,Adami2011}.
\newline\indent
We note that none of the spectroscopically-confirmed XMM-LSS clusters at 0.2 $< z < $ 0.4 \citep{Adami2011} are associated with overdensities in the probability map, even though they have similar 3.6$\micron$ - 4.5$\micron$ colors (see $\S$ 5).  This further highlights the effectiveness of the z$^{\prime}$ - 3.6$\micron$ color cut in eliminating low-redshift interlopers.  The removal of these interlopers also means that the SBS method can be robustly applied to redshifts as low as $z \sim$ 0.8, creating a larger redshift range than can be probed by the \cite{Papovich2008} color cut method.
\begin{figure}
\plotone{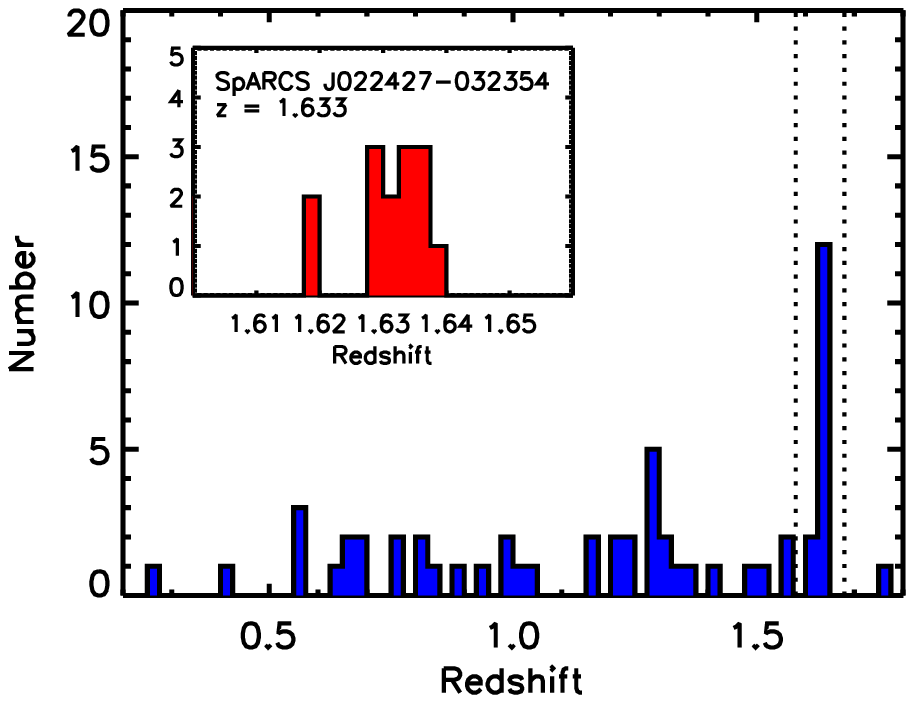}
\caption{\footnotesize Histogram of spectroscopic redshifts in the field of SpARCS J022427-032354.  The inset shows the distribution around the cluster redshift. }
\end{figure}
\begin{figure}
\plotone{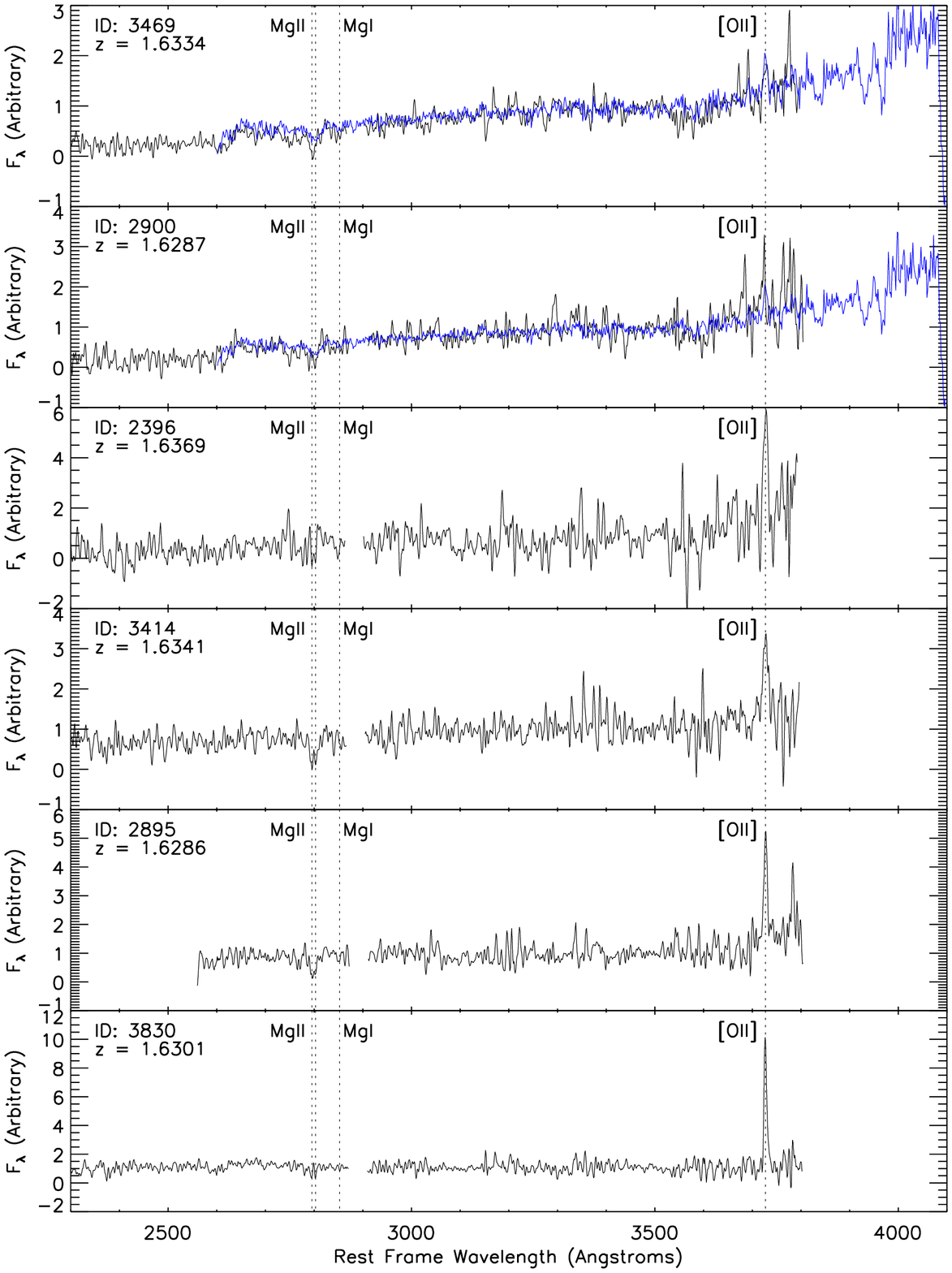}
\caption{\footnotesize Example spectra of cluster members of SpARCS J022427-032354.  Portions of the spectra with strong atmospheric absorption have been masked out.  The locations of the MgII and MgI absorption lines as well as the [OII] emission line are marked.  The majority of the spectroscopic members show strong [OII] emission and MgII absorption suggesting active star formation.  Shown is blue is the mean stacked spectrum of cluster members in a lower redshift cluster that is dominated by galaxies whose star formation is quenched.  The top two spectra show much weaker [OII] emission and have a continuum shape that matches the lower-redshift cluster suggesting they may be quenched galaxies.}
\end{figure}
\section{The Cluster SpARCS J022427-032354 at \lowercase{$z$} = 1.6}
The bottom panel of Figure 4 shows the color slice that is centered at a $z_{photo}$ = 1.6.  There appear to be fewer cluster candidates in the $z =$ 1.6 slice compared to the $z =$ 1 slice.  This is partially because the $z =$ 1.6 map is noisier than the $z =$ 1 map due to the fixed depth of the observations.  Even so, the data are deep enough (M $<$ M$^{*}$ + 1.3) to robustly detect massive clusters so the reduced number of candidates suggests a significant decline in the number of massive clusters between $z = 1.0$ and $z =$ 1.6, consistent with expectations from $\Lambda$CDM models \citep[e.g.,][]{Haiman2001}. 
\newline\indent
The most prominent structure in the $z =$ 1.6 slice is the candidate cluster in the northeast corner of the image which we designate SpARCS J022427-032354.  Also visible in the southwest corner of the field is CIG J0218.3-510 discovered by \cite{Papovich2010}.  That cluster is also one of the more prominent overdensities in the color slice.  
\newline\indent
In the left panel of Figure 5 we show a gz$^{\prime}$3.6$\micron$ color-composite of SpARCS J022427-032354.  The cluster candidate is clearly visible as an overdensity of red galaxies in the center of the image.  Similar to many other high-redshift clusters \citep[e.g.,][]{Eisenhardt2008,Papovich2010,Muzzin2012}, the structure appears to be filamentary without a dominant central galaxy.  
\newline\indent
In the right panel of Figure 3, we have also plotted the 3.6$\micron$ - 4.5$\micron$ colors of galaxies near SpARCS J022427-032354 as a function of 3.6$\micron$.  Similar to the GCLASS clusters, SpARCS J022427-0323354 has a well-defined stellar bump sequence; however, that sequence is substantially redder, consistent with a $z_{photo}$ = 1.6.
Given its high significance and estimated redshift, SpARCS J022427-032354 was designated as a high-priority target for spectroscopic followup in our v1.0 SBS-selected cluster catalog of the XMM-LSS field.  The cluster was observed  using two multi-slit masks with the FORS2 instrument on the VLT as part of Program 085.A-0613(B) (PI: Demarco).  The on-sky integration time for each mask was 3.75 hrs.  The 2D spectra were reduced using standard techniques for flat-fielding, wavelength calibration, and relative flux normalization using the pipeline described in \cite{Demarco2005,Demarco2007}.  Once 1D wavelength-calibrated spectra were extracted, spectroscopic redshifts were determined both by cross correlation and eye-examination of faint spectra for emission lines.  Full details of the procedure will be discussed in a future paper (G. Wilson in preparation).  
\newline\indent
\begin{figure*}
\plotone{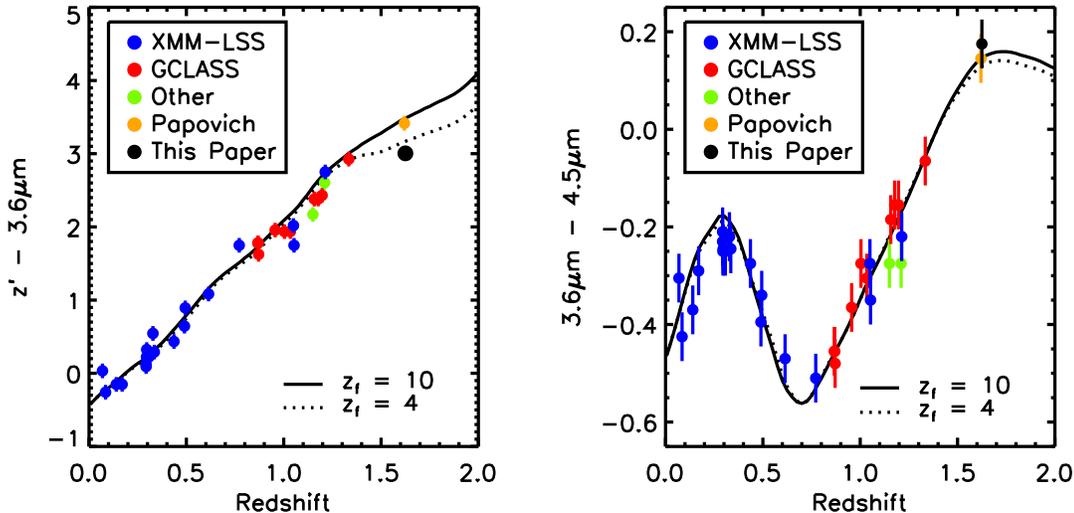}
\caption{\footnotesize Left Panel:  The observed z$^{\prime}$ - 3.6$\micron$ color of the red-sequence as a function of spectroscopic redshift for clusters within the 42 square degree SpARCS survey.   Points are color-coded based on the source of the spectroscopic redshift.  Two SSP models from the BC03 code with formation redshifts of 4.0 and 10.0 are shown and are good descriptions of the data.  Right Panel:  The observed 3.6$\micron$ - 4.5$\micron$ color of the stellar-bump sequence as a function of redshift for the same clusters, with the same BC03 models.  Both the z$^{\prime}$ - 3.6$\micron$ and 3.6$\micron$ - 4.5$\micron$ color-sequences of clusters can be used as an accurate photometric redshift estimate up to $z \sim$ 1.7.}
\end{figure*}
From the two spectroscopic masks a total of 56 high-confidence redshifts were determined.  In Figure 6 we plot the spectroscopic redshift distribution of galaxies within 3 arcmin of SpARCS J022427-032354.  There is a clear overdensity of galaxies at $z =$ 1.63, with a total of 12 galaxies in the range 1.62 $< z <$ 1.64 (see inset of Figure 6).  In Figure 7 several examples of spectra in this redshift range are shown.  The majority of galaxies in this redshift range with spectra show clear [OII] emission (9/12); however, there are three spectra that have weak or possibly no [OII] emission, two of which are shown in Figure 7.  For these galaxies there is a nearby night sky line that makes it difficult to put strong constraints on the level of [OII] emission; however, the continuum shape is similar to that expected in galaxies with quenched star formation.  Overplotted in blue in Figure 7 is the mean stacked spectra of galaxies in the cluster SpARCS J003550-431224 \citep[see][]{Wilson2009,Muzzin2012} which primarily contains quenched galaxies.  The UV continuum of the template matches the UV continuum of the weak [OII] galaxies in Figure 7 quite well, suggesting that these two galaxies may be quenched cluster members at $z = 1.63$.
\newline\indent
In the right panel of Figure 5 the positions of these 12 cluster galaxies are plotted as white squares and the positions of the remaining spectroscopic redshifts as green squares.  Most of the galaxies with redshifts 1.62 $< z <$ 1.64 are concentrated in the central few arcmin of the field suggesting this is a cluster of galaxies as compared to a large, unvirialized "sheet" of galaxies.  There are however several galaxies at the same redshift that extend to $\sim$ 1 Mpc, suggesting that this may be a massive cluster in the process of forming and hence the system may be only partially virialized.  A more detailed analysis of the dynamics of the system will be presented in a future paper (G. Wilson, in preparation).
\newline\indent
We note that there are many red galaxies near the center of the galaxy distribution that do not have redshifts.  Several of these were targeted in the spectroscopic masks but we were unable to determine a spectroscopic redshift due to the target being too faint in the observed optical.  These galaxies have very red z$^{\prime}$ - 3.6$\micron$ colors and could potentially be additional quiescent galaxies in the cluster core.  If they are also at $z = 1.633$, then future NIR spectroscopy would be valuable to determine redshifts using the absorption features that are typically present in galaxies with old stellar populations.  
\newline\indent
Overall, with 12 confirmed members, it appears that SpARCS J022427-032354 is the first confirmed cluster discovered using the SBS selection technique.
\section{Discussion and Conclusions}
The confirmation of SpARCS J022427-032354 shows that the SBS method is capable of selecting high-redshift clusters in combined optical/MIR datasets.  Here we investigate how the SBS photometric redshifts compare to those determined from the typical red-sequence photometric redshifts.
In the left panel of Figure 8 we plot the median z$^{\prime}$ - 3.6$\micron$ colors of red-sequence galaxies from several different samples of clusters in the SpARCS fields as a function of spectroscopic redshift.  The lowest-redshift clusters come from the XMM-LSS cluster sample \citep{Adami2011} and are X-ray selected.  The GCLASS clusters are z$^{\prime}$ - 3.6$\micron$ red-sequence selected.  The ``other" clusters come from \cite{Demarco2010} (red-sequence selection) and \cite{Hashimoto2005} (X-ray selection).  Also shown are CIG J0218.3-510 and SpARCS J022427-032354.  Overplotted are two single-burst models with formation redshifts of $z_{f}$ = 4.0 and $z_{f}$ = 10.0.  The models are good descriptions of the redshift evolution of the color of the observed red-sequence and the rms scatter of the data compared to the model is 0.18 mag, implying that the z$^{\prime}$ - 3.6$\micron$ red-sequence color provides photometric redshifts for clusters to an accuracy of $\delta$$z$ = 0.08 from 0.1 $< z <$ 1.6.  
\newline\indent
In the right panel of Figure 8 we plot the median 3.6$\micron$ - 4.5$\micron$ colors of the stellar bump sequence for the same clusters, as well as the same single-burst models.  The model is also a good description of those data.  The rms scatter in color is 0.05, which implies a redshift accuracy of $\delta$$z$ = 0.05 for clusters at 0.8 $< z <$ 1.6.  This is better than the z$^{\prime}$ - 3.6$\micron$ colors as a photometric redshift and is surprising given that 3.6$\micron$ - 4.5$\micron$ spans a much smaller range over this redshift.  The slightly better $z_{photo}$ may be because it is less sensitive to star-formation history, or because the MIR zeropoints are more stable than ground-based zeropoints over wide fields.  Regardless, Figure 6 demonstrates just how well the models describe the color evolution of the population and that good quality $z_{photo}$ can be determined using the 3.6$\micron$ - 4.5$\micron$ colors of galaxies in candidate clusters.  \newline\indent
One caveat with the SBS method is that it is likely to have a higher false-positive detection rate compared to other optical cluster detection techniques such as the red-sequence method.  Although the method does use 3.6$\micron$ - 4.5$\micron$ color slices, the observed 3.6$\micron$ - 4.5$\micron$ colors of galaxies do not evolve as rapidly with redshift as observed colors that span the 4000\AA~break such as z$^{\prime}$ - 3.6$\micron$.  Therefore, line-of-sight projection will occur more frequently.  The false-positive rate for red-sequence selection is $\sim$ 5\% \citep[see,][]{Gladders2005,Gilbank2007}.  Given the limited color range, the SBS false-positive rate may be as much as several times higher than this.  At present, with only one confirmed cluster it is not possible to quantify the purity level of cluster samples selected with SBS.  In the future, quantifying the false-positive rate will be important for using large samples of these clusters.  
\newline\indent
Overall, the SBS algorithm appears to be an effective means of identifying candidate high-redshift clusters in combined optical/MIR data.  It is based on combining the cluster red-sequence method of \cite{Gladders2000} with the MIR method of \cite{Papovich2008}; however, the combination of the two provides distinct advantages over each individually.  Using the MIR colors of galaxies allows the cluster searches to continue to higher redshift over wider areas than is feasible using ground-based optical imaging.  Likewise, the use of MIR color slices and the optical/MIR color cut reduces the low-redshift interloper fraction and the probability of line-of-sight projections which increases the purity level of the cluster sample.  As shown in Figure 8, it also allows for an estimate of the photometric redshift for all clusters in the survey with good accuracy.  
\newline\indent
The SBS algorithm has potential to provide a large, homogenously-selected sample of clusters in the coming years.  We have already applied it to the 50 degree$^2$ SWIRE fields and discovered several hundred cluster candidates at $z >$ 1.  In this paper we presented confirmation of the cluster SpARCS J022427-032354 at $z =$ 1.63; however, several other clusters discovered with the method in the SWIRE area have also been confirmed and will be discussed in future papers (G. Wilson, in preparation, R. Demarco, in preparation).  In the future the method could also be applied to the available wide-field imaging datasets from $Spitzer$.  At present, these cover of order several hundred square degrees and could provide an intermediate-mass cluster sample up to $z \sim$ 1.7, something which has potential for providing both interesting constraints on the growth of massive structures as well as the evolution of the cluster galaxy population.  
\acknowledgements
RD gratefully acknowledges the support provided by the BASAL Center for Astrophysics and Associated Technologies (CATA), and by FONDECYT grant N. 1100540.

\bibliographystyle{apj}
\bibliography{apj-jour,myrefs}




\end{document}